\documentclass[aps,pra,floatfix,twocolumn,noshowpacs,showkeys,nofootinbib]{revtex4-1}
\usepackage{xcolor, graphicx, ulem}
\usepackage{amsmath, amssymb}
\usepackage[colorlinks=true,urlcolor=blue,citecolor=blue,linkcolor=blue]{hyperref}

\begin{document}

\title{Accessing the non-equal-time commutators of a trapped ion}

\author{F. Krumm}\email{fabian.krumm@uni-rostock.de}\affiliation{Arbeitsgruppe Theoretische Quantenoptik, Institut f\"ur Physik, Universit\"at Rostock, D-18059 Rostock, Germany}
\author{W. Vogel}\affiliation{Arbeitsgruppe Theoretische Quantenoptik, Institut f\"ur Physik, Universit\"at Rostock, D-18059 Rostock, Germany}

\begin{abstract} 
The vibronic dynamics of a trapped ion in the resolved-sideband regime can be described by the explicitly time-dependent nonlinear Jaynes-Cummings model.
It is shown that the expectation value of the interaction Hamiltonian and its non-equal-time commutator can be determined by measuring the electronic-state evolution.
This yields direct insight into the time-ordering contributions to the unitary time evolution.
In order to prove extraction of the quantities of interest works for possibly real data, we demonstrate the procedure by means of generated data.
\end{abstract}

\date{\today}
\maketitle

\section{Introduction} 
Starting with the development of quantum mechanics and the introduction of Hilbert-space operators, the noncommutativity of the latter became an issue.
It leads to many fascinating physical effects, where the most prominent example is most likely the Heisenberg uncertainty principle~\cite{Lit:heis1,Lit:heis2,Lit:heis3}.
Furthermore, non-commutativity plays an important role in quantum field theory~\cite{Lit:fieldtheory}, quantum many-body systems~\cite{Lit:manybody1,Lit:manybody2,Lit:manybody3,Lit:manybody4,Lit:manybody5,Lit:manybody6,Lit:manybody7,Lit:manybody8,Lit:manybody9}, quantum electrodynamics~\cite{Lit:qed1,Lit:qed2,Lit:qed3}, the standard model~\cite{Lit:standmod}, and cosmology~\cite{Lit:cos1,Lit:cos2}.
Here we consider the problem of non-equal-time commutators from the quantum optics point of view.

A noteworthy achievement in this context is the experimental verification of the bosonic commutation relation, $[\hat a,\hat a^\dag]=\hat 1$.
Although this relation is of fundamental relevance for the formulation of quantum mechanics, it was not verified before 2007, in a seminal paper by Bellini and co-authors~\cite{Lit:P07}.
Later on, this subject was analyzed in some more detail~\cite{Lit:Kim08,Lit:Zav09}.
Elementary commutation rules of such a type are equal-time rules introduced in the procedure of canonical quantization.

This leads to another fundamental subject, namely, the non-equal-time commutation rules, which play an important role in the context of interaction problems including \textit{time ordering}.
If the dynamics of an explicitly time-dependent Hamiltonian is formally solved in terms of the standard time-evolution operator, one finds that the latter obeys a time-ordering prescription (cf., e.g.,~\cite{Lit:Schleich,Lit:VW06,Lit:GrynbergAF2010,Lit:Agarwal2013}).
This prescription must not be omitted as it has a crucial impact on the dynamics of the system~\cite{Lit:KVW87,Lit:Cresser87,Lit:KVW90,Lit:ChristBMS2013,Lit:QuesadaS2014,Lit:QuesadaS2015,Lit:QuesadaS2016,Lit:KrummSV2016}.
Paradoxically, despite its key role in basic quantum mechanics, detailed treatments of time-ordering effects are rarely available.
A direct verification of the non-equal-time commutators of Hamiltonians has, to our best knowledge, not been studied  yet. 
Of course, the non-equal-time commutators of interest only occur in the case of explicitly time-dependent Hamiltonians. 
We also stress that the time-dependent commutators are not postulated in the quantization procedure. 
Instead, they require the solution of the interaction problem under consideration. 
Hence, it is very useful to consider an exactly solvable interaction dynamics. As the latter should also not be a trivial example, we consider the nonlinear vibronic interaction of a trapped and laser-driven ion. For a slightly off-resonant driving laser, we are just in the regime of interest.

In this work, we use basic relations of quantum mechanics to show that the measurement of the expectation value of an explicitly time-dependent interaction Hamiltonian yields the expectation value of a partly integrated non-equal-time commutator of this Hamiltonian.
If this commutator is nonzero, the system undergoes a time-ordered dynamics.
In principle, the latter can be determined for any physical system with an explicitly time-dependent Hamiltonian. 
For a rigorous treatment of the problem, we focus on the mentioned exactly solvable problem. 
Note that insight into the non-equal-time commutators is an issue of relevance for the general dynamics of quantum systems. 
In many cases, when exact solutions are not available, the problem can only by solved numerically. 
For the trapped-ion dynamics under study, the advantage is that we may obtain the expectation value of the interaction Hamiltonian directly from the measurement of the excited electronic-state occupation probability.
The specific steps of the procedure will be demonstrated by the use of generated data.

The paper is structured as follows. 
In Sec.~II we introduce the time evolution in the case of explicitly time-dependent interaction Hamiltonians together with the resulting non-equal-time commutators. 
The nonlinear explicitly time-dependent Jaynes-Cummings model is introduced in Sec.~III, which allows us to study the dynamics of interest on the basis of exact solutions. 
In Sec.~IV we show how one may experimentally determine the interaction Hamiltonian in Fock basis for the case of the laser-driven zeroth motional sideband of the ion.
Section V is devoted to the investigation of the relevance of the non-equal-time commutators of the interaction Hamiltonian. 
A summary and some conclusions are given in Sec.~VI.

\section{Time evolution}
	We start with some fundamental relations of quantum theory.
	The properties of a physical system may be compactly expressed by its Hamiltonian, $\hat H_S(t)=\hat H_{0,S}+ \hat  H_{\text{int},S}$.
	The index $S$ denotes the Schr\"odinger picture, $\hat H_{0,S}$ is the free evolution of the system, and $\hat  H_{\text{int},S}$ is the interaction of different degrees of freedom. 
	In the interaction picture, denoted by the index $I$, and assuming that the interaction Hamiltonian is in this picture explicitly time dependent, the dynamics of the system is  described by the time-evolution operator
	\begin{align}
	    \hat U_I(t) = \mathcal T \exp \left( - \frac{i}{\hbar} \int^t_{0} \hat H_{\text{int},I} ( \tau) d\tau \right).
	\end{align}
	Throughout this work, the explicit time dependence of the Hamiltonians is presumed.
	Here, $\mathcal T$ denotes the time-ordering prescription which only can be ignored if the interaction Hamiltonian commutes with itself at different times, 
	$[\hat H_{\text{int},I} ( \tau_1),\hat H_{\text{int},I} ( \tau_2)]=0, \ \forall (\tau_1,\tau_2)$
	(see, e.g.,~\cite{Lit:Schleich,Lit:VW06,Lit:GrynbergAF2010,Lit:Agarwal2013}).
	
	We emphasize that throughout this work the time dependence of the Hamiltonian $ \hat H_{\text{int},I} ( \tau)$ refers to the explicit time dependence and not to the (implicit) time dependence of the operators.
	The latter is directly caused by the time-evolution operator $\hat U_I(t)$.
	In general, the interaction Hamiltonian is proportional to some coupling constant $|\kappa|$ and, hence, we may use a power series expansion
	\begin{align}
	  &\hat U_I(t) 
	  = 1 - \frac{i}{\hbar} \int_0^t d{\tau_1} \hat H_{\text{int},I}(\tau_1)  \nonumber \\
	  &- \frac{1}{\hbar^2} \int_0^t d{\tau_1}  \int_0^{\tau_1}  d{\tau_2}  \hat H_{\text{int},I}(\tau_1)  \hat H_{\text{int},I}(\tau_2) +  O(|\kappa|^3).
	\end{align}
	The full time evolution of the interaction Hamiltonian reads as,
	\begin{align}
	\label{Eq:expansionK}
	&\hat U_I^\dag (t)  \hat H_{\text{int},I}(t) \hat U_I(t) \\
	&= \underbrace{ \hat H_{\text{int},I}(t)}_{\propto |\kappa|}   + \frac{i}{ \hbar} \int_{0}^t d\tau_1 \underbrace{ \left[  \hat H_{\text{int},I}  (\tau_1) , \hat H_{\text{int},I}(t) \right]}_{\propto |\kappa|^2} + O(|\kappa|^3 ). \nonumber
	\end{align}
	The terms proportional to $|\kappa|$ and $|\kappa|^2$ yield the interaction Hamiltonian and its partly integrated non-equal-time commutator, respectively.
	However, especially from the experimental point of view, the determination of the expectation value of solely the interaction Hamiltonian is not a trivial task.
	In the following, we will consider a realistic model, the explicitly time-dependent nonlinear Jaynes-Cummings Hamiltonian, which describes the vibronic dynamics of a trapped ion in the resolved-sideband regime.
	We will show that for this model the expectation value of Eq.~\eqref{Eq:expansionK} can be derived from an experimentally accessible observable.

\section{Nonlinear Jaynes-Cummings model} 
	The quantized center-of-mass motion of a trapped ion, in the resolved-sideband limit, can be described by the nonlinear Jaynes-Cummings model~\cite{Lit:VF95}.
	Including a frequency mismatch $\Delta \omega$, which we assume to be small but  nonzero, such that the Hamiltonian is explicitly time dependent in the interaction picture.
	The corresponding $k$th-order nonlinear interaction Hamiltonian, after a vibrational rotating wave approximation, reads as
	\begin{align}
		\label{Eq:Hinteraction}
		\hat H_{\text{int},I}(t)= \hbar |\kappa| e^{- i\Delta \omega t + i \theta} \hat A_{21}  \hat f_k(\hat n;\eta) \hat a^k+ \text{H.c.}
	\end{align}
	(see Ref.~\cite{Lit:K18} for a detailed derivation).
	Here, $\kappa=|\kappa|e^{i \theta}$ is the coupling constant of the ion's electronic and vibrational levels and is proportional to the amplitude of the driving laser.
	Additionally, $\hat a$  and $\hat a^\dag$ are the annihilation and creation operators of the vibrational mode and, in the case of a standing wave, with $\hat n = \hat a^\dag \hat a$, $\hat f_k(\hat n;\eta)$ 
	describes the mode structure of the driving laser field at the position of the ion.
	It is, in Fock basis, defined as follows:
	\begin{align}
		& \hat f_k(\hat n;\eta)   \\
		&=\frac{1}{2} e^{i \Delta \phi -\eta^2/2}  \sum_{n=0}^\infty |n\rangle \langle n | \frac{(i \eta)^k n!}{(n+k)!} L_n^{(k)} (\eta^2) + \text{H.c.}  \nonumber , 
	\end{align}
	with $L_n^{(k)} $ denoting the generalized Laguerre polynomials,
	$\eta$ is the Lamb-Dicke parameter, and $\Delta \phi$ determines the position of the trap potential relative to the laser wave. 
	The atomic flip operator $\hat A_{ij}=|i\rangle \langle j|$ ($i,j=1,2$) describes the $|j\rangle \rightarrow |i\rangle$ transition.
	Furthermore, the classical driving laser with frequency $ \omega_L=\omega_{21}-k \nu + \Delta \omega$ is slightly detuned from the $k$th sideband by $\Delta \omega$, which yields the time dependence of the Hamiltonian in Eq.~\eqref{Eq:Hinteraction}.
	 Here, $\nu$ is the trap frequency and $\omega_{21}=\omega_2 - \omega_1$ is the separation of the electronic levels $|1\rangle$ and $|2 \rangle$.
	 Finally, the Hamiltonian describing the free evolution reads as
	 \begin{align}
	  \hat H_{0,I}= \hbar \nu \hat n + \hbar \omega_{21}\hat A_{22}.
	 \end{align}
	A detailed discussion of the Hamiltonians can be found in Refs.~\cite{Lit:VF95,Lit:K18} or Chap. 13 of~\cite{Lit:VW06}.
	
	The solution of the corresponding dynamics, 
	\begin{align}
		\label{Eq:solutionU}
		&\hat U_I(t) =\sum_{n=0}^\infty \Big( a_n(t) |2, n \rangle \langle 2,n| \\
		& - b_n^\ast(t) e^{-2i \theta} |1,n+k \rangle \langle 2,n|  + b_n(t) e^{2i \theta} |2,n\rangle \langle 1,n+k|   \nonumber \\
		& + a_n^\ast(t) |1,n+k \rangle \langle 1,n+k| \Big) +\sum\limits_{q=0}^{k-1}|1,q\rangle\langle 1,q|, \nonumber
	\end{align}
	with
	\begin{align}
		&a_n( t )=e^{-i \Delta\omega  t /2}
		\bigg[
		\cos(\Gamma_n t )
		+ \frac{i \Delta\omega}{2 \Gamma_n}\sin(\Gamma_n t)
		\bigg],
	 \\
		&b_n( t )=e^{-i \Delta\omega t /2}\frac{|\kappa| w_n}{i \Gamma_n}
		\sin(\Gamma_n t )
		, \nonumber  \\
		& \Gamma_n=\sqrt{\left(\frac{\Delta\omega}{2}\right)^2+w_n^2|\kappa|^2}, \nonumber \\
		&w_n= \cos\left(\Delta\phi+\frac{\pi}{2}k\right)\eta^k e^{-\eta^2/2}\sqrt{\frac{n!}{(n+k)!}}L^{(k)}_n (\eta^2) \nonumber
	\end{align}
	  has been derived in Ref.~\cite{Lit:Tobi}.
	
	Let us consider the time evolution of the occupation probability of the excited electronic state, $\sigma_{22}=\langle \hat A_{22} \rangle$.
	Because of $[\hat A_{22} ,\hat H_{0,I}] =0$, $\sigma_{22}$ depends solely on the interaction Hamiltonian, which yields
	\begin{align}
	  &\dot \sigma_{22}(t)=\frac{i}{\hbar} \left\langle  \hat U_I^\dag(t) \left[ \hat H_{\text{int},I}(t), \hat A_{22} \right] \hat U_I(t) \right\rangle .
	\end{align}
	Using the Hamiltonian in Eq.~\eqref{Eq:Hinteraction}, we obtain
	 \begin{align}
	 \label{Eq:sigma22-a}
	 &  \dot \sigma_{22}(t) =  i |\kappa| \left\langle - e^{- i\Delta \omega t + i \theta} \hat A_{21}(t)  \hat f_k(\hat n(t);\eta) \hat a^k(t) \right.\\\nonumber 
	  &\left.  +e^{ i\Delta \omega t - i \theta} \hat A_{12}(t)   \hat a^{\dag k}(t) \hat f_k(\hat n(t);\eta)  \right\rangle .
	\end{align}
	Comparing this expression with the Hamiltonian~\eqref{Eq:Hinteraction}, for $\Delta \omega \neq 0$ we get
	\begin{align}
	\hbar \Delta \omega \dot \sigma_{22}(t) \equiv  \left\langle \hat U_I^\dag(t) \left( \frac{d}{dt}  {\hat H}_{\text{int},I}(t)  \right)\hat U_I(t) \right\rangle .
	\end{align}
	Note that for $\Delta \omega=0$, when the Hamiltonian is not explicitly time dependent, both sides of the latter equation vanish and hence they yield no physical insight in the interaction dynamics.
	Since $\hat U_I^\dag(t) [ \frac{d}{dt}\hat H_{\text{int},I}(t) ]  \hat U_I(t)  = \frac{d}{dt} [ \hat U_I^\dag(t) \hat H_{\text{int},I}(t) \hat U_I(t)  ] $, we may integrate Eq.~\eqref{Eq:sigma22-a} to arrive at
	\begin{align}
	 \label{Eq:sigma22-b}
	   &  \hbar \Delta \omega \left[\sigma_{22}(t) - \sigma_{22}(0) \right] \\ \nonumber 
	 & = \left\langle \hat U_I^\dag(t) {\hat H}_{\text{int},I}(t) \hat U_I(t) \right\rangle  -  \left\langle \hat U_I^\dag(0) {\hat H}_{\text{int},I}(0) \hat U_I(0) \right\rangle . 
	\end{align}

	We observe that the measurement of the excited-state occupation probability $\sigma_{22}(t)$, which is achieved via probing an auxiliary transition for resonance fluorescence~\cite{Lit:N86,Lit:S86,Lit:B86}, is directly related to the expectation 
	value of the time-dependent interaction Hamiltonian.
	The consideration of different orders with respect to $|\kappa|$ allows one to determine either the interaction Hamiltonian itself or the corresponding commutator in Eq.~\eqref{Eq:expansionK}.
	Without loss of generality, we set $\theta = 0$ in the following.
	
	In this section, we have briefly recapitulated the detuned nonlinear Jaynes-Cummings model which describes the quantized motion of a trapped ion in the resolved-sideband regime.
	This model was originally introduced for zero detuning~\cite{Lit:VF95} and experimentally proven to properly describe the experimental dynamics of trapped ions~\cite{Lit:Meek96}. 
	In experiments the extension of the model to include the detuning under study here is a minor issue.
	
	For the case of detuning we have shown that the expectation value of the interaction Hamiltonian can be obtained from the occupation probability $\sigma_{22}(t)$ of the (excited) electronic state.
	According to Eq.~\eqref{Eq:expansionK}, from the latter we can extract the expectation value of the interaction Hamiltonian in the interaction picture ($\propto |\kappa|$) and the corresponding commutator ($\propto |\kappa|^2$).
	In the next two sections we will demonstrate the procedure step by step by using generated data.
	The latter are used to visualize the situation for experimental, i.e. fluctuating, data.
	
\section{Determination of the zeroth sideband interaction Hamiltonian}

	In this section we will consider the determination of the interaction Hamiltonian in Fock basis, 
	$\langle \hat H_{\text{int},I}(t)  \rangle = \ \text{Tr}_\text{el} [ \hat \sigma(0) \langle n | \hat H_{\text{int},I}(t) | n \rangle ]$, for $k=0$ in Eq.~\eqref{Eq:Hinteraction}. 
	Note that this Hamiltonian is diagonal in Fock basis.
	A remark concerning the case $k>0$ is given at the end of this section.
	Here, $\text{Tr}_\text{el} $ is the trace over the electronic degrees of freedom.
	The generation of vibrational Fock states in an ion trap was already investigated in the 1990s (cf. Refs.~\cite{Lit:Meek96,Lit:V96}).
	
	In the following we will use the input density matrix $\hat \rho(0)= \hat \sigma(0) \otimes \hat \rho_\text{mot}(0) $, where $\hat \sigma(0)$ and $ \hat \rho_\text{mot}(0)$ describe the electronic and the motional input state, respectively.
	An overview over experimentally possible states of a trapped ion can be found in Ref.~\cite{Lit:K18}, and references therein.
	If the electronic state is initially in a superposition,
	\begin{align}
		\label{Eq:secivelectronic}
		\hat \sigma (0) = ( \gamma_1 |1 \rangle +\gamma_2 |2 \rangle ) ( \gamma_1^\ast \langle 1| + \gamma_2^\ast \langle 2|),
	\end{align}
	with $|\gamma_1|^2+|\gamma_2|^2=1$, and $\hat \rho_\text{mot} = | n\rangle \langle n|$,
	one readily derives
	\begin{align}
	      \label{Eq:expecH0}
		 \left\langle \hat U_I^\dag(0) {\hat H}_{\text{int},I}(0) \hat U_I(0) \right\rangle= \hbar |\kappa|  f_0(n;\eta) ( \gamma_1 \gamma_2^\ast + \gamma_2 \gamma_1^\ast).
	\end{align}
	Here we defined $f_k(n;\eta)=\langle n | \hat f_k(\hat n;\eta) |n\rangle$.
	Hence, if  $\text{arg}(\gamma_1) -\text{arg}(\gamma_2)= (2m+1) \frac{\pi}{2}$ for $m=0,1,\dots$, then the expectation value in Eq.~\eqref{Eq:expecH0} becomes zero.
	Thus, we set $\gamma_1 = e^{i \pi/2} /\sqrt{2}$ and $\gamma_2=1/\sqrt{2}$, which leads to $\sigma_{22}(0)=1/2$.
	Hence, Eq.~\eqref{Eq:sigma22-b} simplifies to
	\begin{align}
		\label{Eq:postsimudisc}
		 \left\langle \hat U_I^\dag(t) {\hat H}_{\text{int},I}(t) \hat U_I(t) \right\rangle=\hbar \Delta \omega [ \sigma_{22}(t) -1/2 ] .
	\end{align}
	
	To demonstrate that our approach applies to experimental data, we will generate random numbers which approximate the distribution which $\sigma_{22}$ obeys.\footnote{For the generation of random numbers from a given distribution the \textit{Mathematica} inbuilt method \textsc{RandomVariate} was used.}
	By using this approximated distribution, a sequence of artificial data points is obtained which statistically fluctuate around the exact evolution of $\sigma_{22}$.
	For convenience, we introduce the dimensionless coupling $g$, i.e., a rescaling, via $|\kappa| \rightarrow g  |\kappa'|$ and the dimensionless time $|\kappa'| t$.
	
\begin{figure}[h]
	\centering
	\includegraphics*[width=8.6cm]{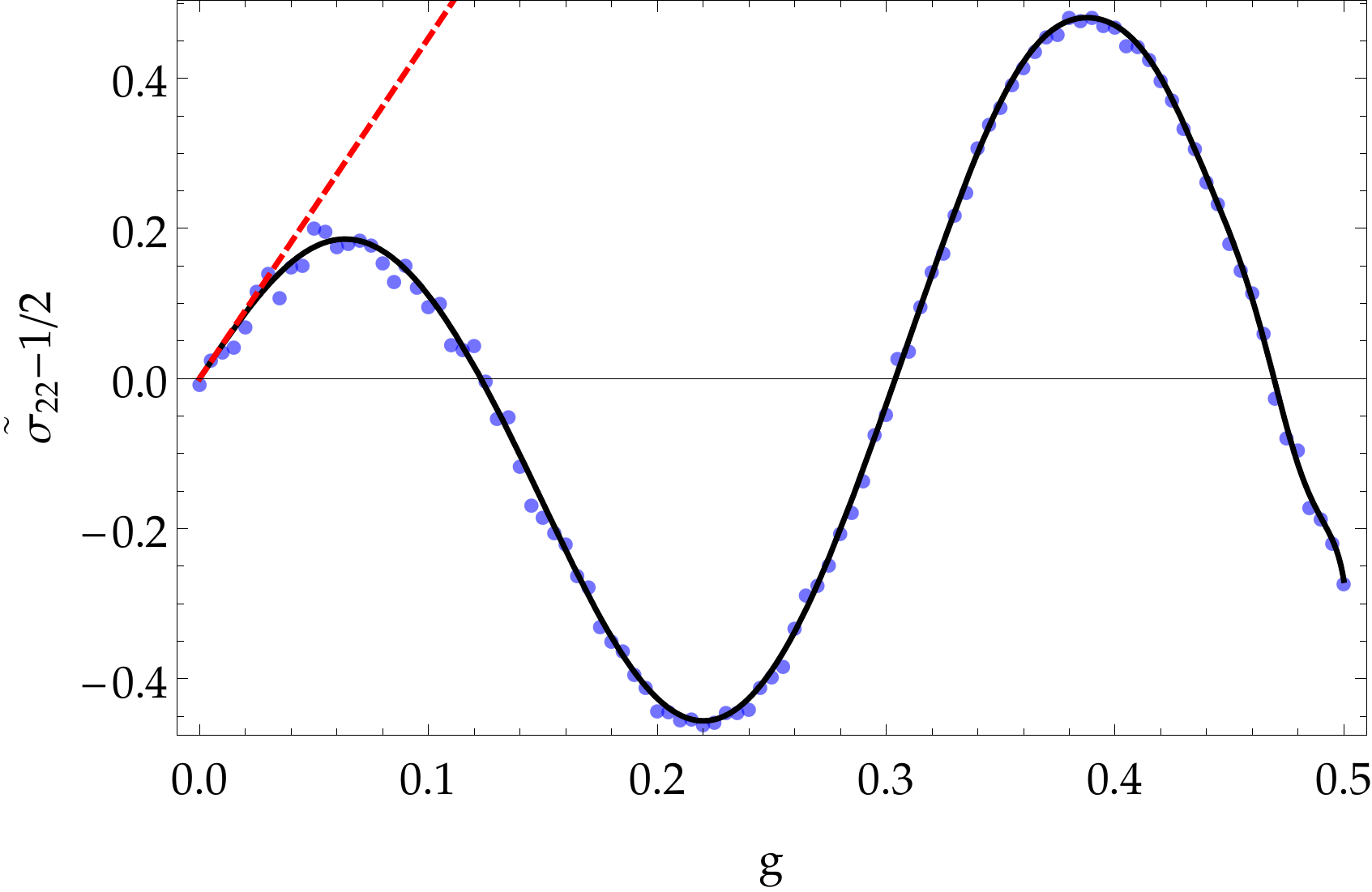}
	\caption{The generated data (blue dots) of the excited state occupation probability together with a nonlinear curve fit [Eq.~\eqref{Eq:fit1}] for the excitation to the zeroth sideband, $k=0$, at $|\kappa'| t = 10$ (solid black line).
		The motional input state is the ground state $|  n = 0 \rangle$.
		The quantity $c_1 g$ is given as the dashed red line.
	 	 Parameters: $\eta=0.2$, $\Delta \omega/|\kappa'|=0.2$, $\Delta \Phi=0$, and $\nu=5000$.}\label{Fig:new-final1}
	\end{figure}

	A first result of the basic procedure is shown in Fig.~\ref{Fig:new-final1} for a fixed time.
	Therein, each value of $\sigma_{22}$ (blue dots) is obtained from $10^3$ random numbers, to mimic the distribution of $\sigma_{22}$ for a fixed $n$ for the motional input state $| n\rangle \langle n|$.
	They are fitted by the polynomial
	\begin{align}
		\label{Eq:fit1}
		\tilde \sigma_{22} - \frac{1}{2} = \sum_{l\geq 0} c_{2l+1}g^{2l+1}.
	\end{align} 
	In the fit function only odd orders of $g$ appear, due to the structure of the Hamiltonian~\eqref{Eq:Hinteraction}---in this case due to the algebra of the atomic flip operators---and our choice of the electronic input state.
	The parameter $c_1$ leads to the desired Hamiltonian [cf. Eq.~\eqref{Eq:expansionK}] and is visualized in Fig.~\ref{Fig:new-final1} via the dashed red line.
	It is obvious that especially at $g \ll 1$ a meticulous resolution of the data is important. Here we note that in experiments the dependence on the coupling strength $g$, as considered in Fig.~\ref{Fig:new-final1}, can be well controlled through the amplitude of the laser driving the trapped ion. For details on this dependence, we refer to Sec. 13.3 of Ref. [24].
	
\begin{figure}[b]
	\centering
	\includegraphics*[width=8.6cm]{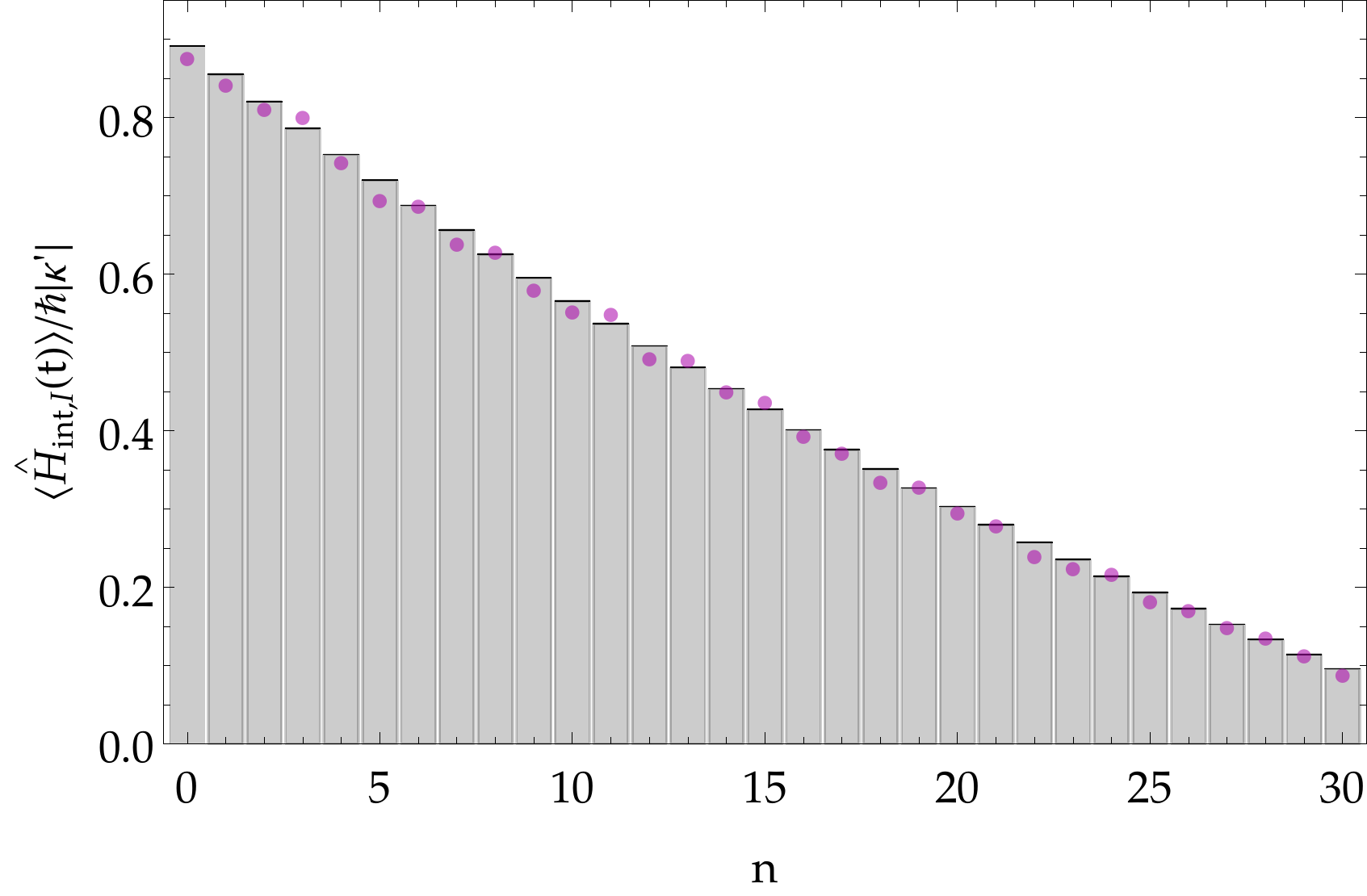}
	\caption{Generated data (magenta dots) obtained by the technique in Fig.~\ref{Fig:new-final1}, for motional Fock states $| n\rangle$ at $|\kappa'| t = 10$. 
	The other parameters are the same as in Fig.~\ref{Fig:new-final1}. 
	The gray bars represent the analytical results according to Eq.~\eqref{Eq:analyticalH}. }\label{Fig:new-final2}
	\end{figure}

	Repeating this procedure for various Fock input states $ | n \rangle$ yields the interaction Hamiltonian in Fock-space representation (see Fig.~\ref{Fig:new-final2}).
	Here we increased the number of random events to $5 \times 10^3$.
	The theoretical prediction (gray bars) of the expectation value of the interaction Hamiltonian in the Fock state $|n\rangle$ is easily calculated to be
	\begin{align}
	\label{Eq:analyticalH}
		 \langle  \hat H_{\text{int},I}(t)   \rangle = \hbar |\kappa| f_0(n;\eta) \left ( \gamma_1 \gamma_2^\ast e^{- i \Delta \omega t} + \gamma_2 \gamma_1^\ast e^{i \Delta \omega t} \right).
	\end{align}
	On this basis we easily obtain, for the case under study, the expectation value for an arbitrary motional quantum state [$\hat \rho_\text{mot}(0)$] as
	\begin{align}
	 & \text{Tr}\left\{  \left[ \hat \rho_\text{mot} (0) \otimes \hat \sigma (0) \right]   \hat H_{\text{int},I}(t)  \right\}  \\ 
& =    \hbar |\kappa|  \left( \gamma_1 \gamma_2^\ast e^{- i \Delta \omega t} + \text{c.c.} \right) \sum_{n=0}^\infty P_n f_0(n;\eta)  ,  \nonumber
	\end{align}
           in which $\hat \sigma(0)$ is given in Eq.~\eqref{Eq:secivelectronic} and $P_n$ is the  number statistics of the motional quantum state under consideration.
	The depicted results, which were derived from the generated data, are close to the analytical results.
	In certain situations the extraction of the expectation value of the interaction Hamiltonian could also serve as a consistency check before investigating the non-equal-time commutators, which will be considered in the next section.
	Hamiltonians which are not diagonal in the Fock basis can be accessed via its determination in the coherent state basis. The  subsequent integration over the Glauber-Sudarshan $P$~function yields the expectation values of the more general interaction Hamiltonians (for $k \neq  0$) in Eq.~\eqref{Eq:Hinteraction} (see, e.g., Ref.~\cite{Lit:VW06}).

\section{Accessing the commutator} 

	For this task we use $\hat \sigma(0) = |1 \rangle \langle 1|$. 
	Hence, the ion is initially in the electronic ground state, so that $\sigma_{22}(0)=0$.
	From Eq.~\eqref{Eq:sigma22-b} we get
	\begin{align}
	\left\langle \hat U_I^\dag(t) {\hat H}_{\text{int},I}(t) \hat U_I(t) \right\rangle = \hbar \Delta \omega \sigma_{22}(t) .
	\end{align}
	Furthermore, we assume that the vibrational input state is a coherent state $|\alpha_0\rangle$.
	Details concerning the preparation of coherent motional states can be found, for example, in Ref.~\cite{Lit:Meek96}.

\begin{figure}[t]
	\centering
	\includegraphics*[width=8.6cm]{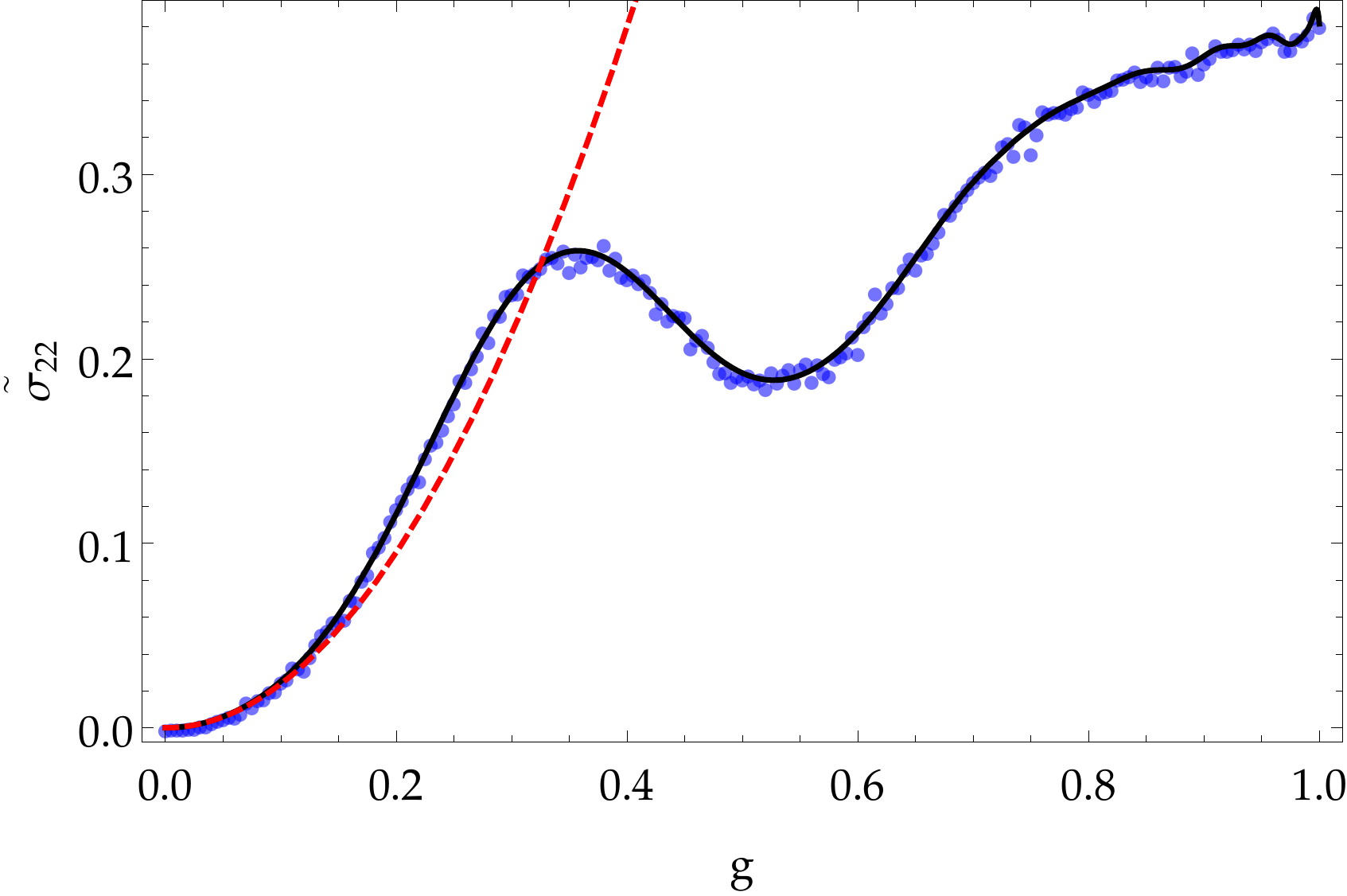}
	\caption{The generated data (blue dots) of the excited state occupation probability together with a nonlinear curve fit $\tilde \sigma_{22}$ [Eq.~\eqref{Eq:fit}] for the excitation to the second sideband, $k=2$, are shown for $|\kappa'| t = 40$ (solid black line).
		The quantity $c_2 g^2$ is given as the dashed red line.
		Parameters: $\alpha_0=\sqrt{12}$, $\eta=0.2$, $\Delta \omega/|\kappa'|=0.2$, $\Delta \Phi=0$, and $\nu=5000$.
	}\label{Fig:final11}
	\end{figure}

	In Fig.~\ref{Fig:final11} we outline the basic procedure, where the statistics is approximated by using $10^4$ random numbers for each data point; for explanations see the discussion following Eq.~\eqref{Eq:postsimudisc}.
	The generated data are now fitted by the function
	\begin{align}
		\label{Eq:fit}
		\tilde \sigma_{22} = \sum_{l\geq1} c_{2l}g^{2l}.
	\end{align}
	For similar reasons as in Eq.~\eqref{Eq:fit1} now only even orders of $g$ appear. 
	According to Eq.~\eqref{Eq:expansionK}, the parameter $c_2$ yields the desired time-integrated commutator in Eq.~\eqref{Eq:expansionK}.
	The parameter $c_2$ is visualized in Fig.~\ref{Fig:final11} by the dashed red line and describes the quadratic contribution which represents the sought commutator.
	\begin{figure}[b]
	\centering
	\includegraphics*[width=8.6cm]{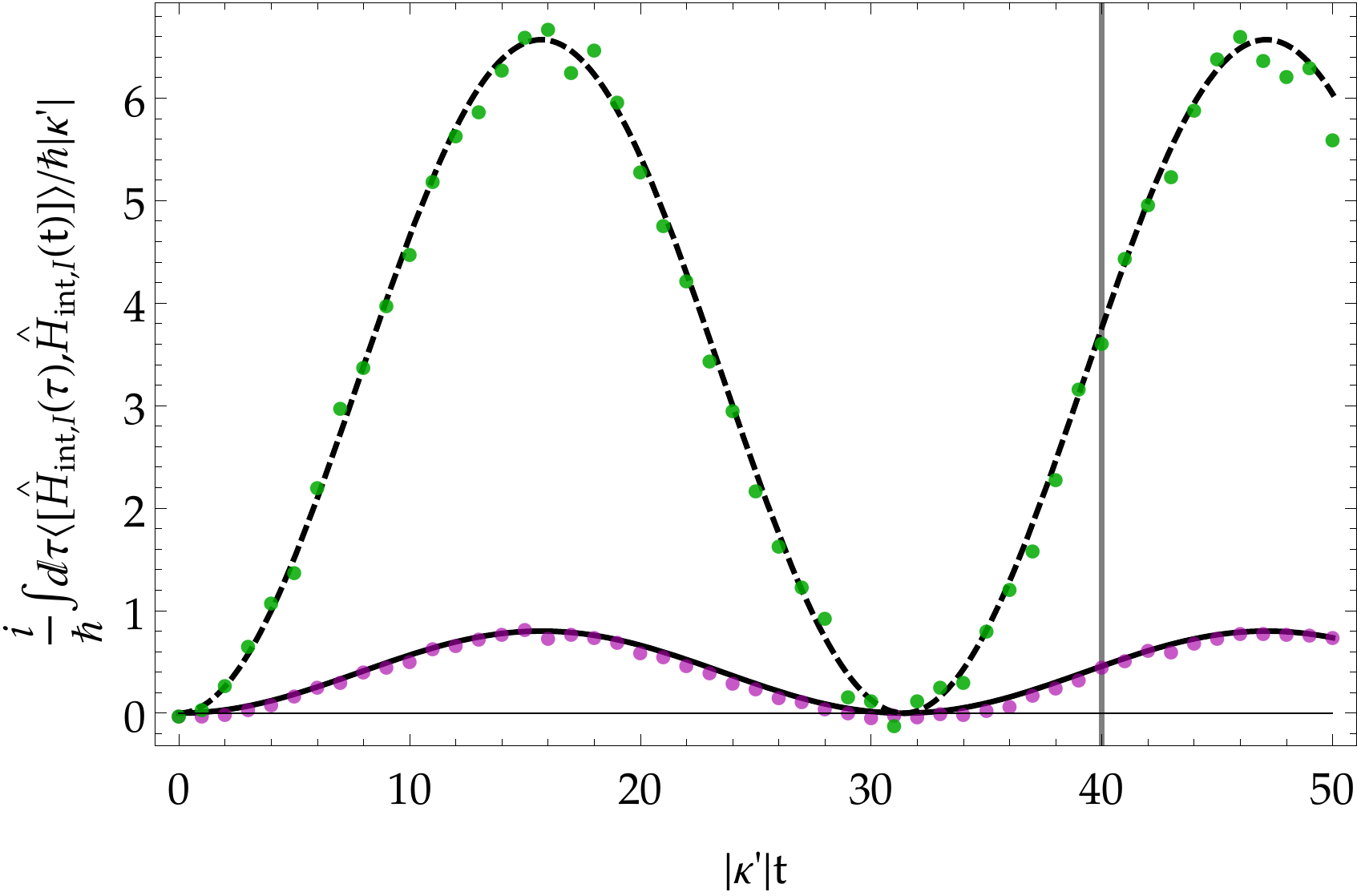}
	\caption{The generated data together with the theoretical predictions of the expectation value of the time-integrated commutator (black lines) from Eq.~\eqref{Eq:JCcommutator}.
		 The data correspond to the scenarios $k=2$ (magenta dots) and $k=0$ (green dots).
		The gray line at $|\kappa'|t =40$ marks the situation depicted in Fig.~\ref{Fig:final11} for the $k=2$ case.
		The other parameters are the same as in Fig.~\ref{Fig:final11}.}\label{Fig:final22}
	\end{figure}	

	To finally obtain the time evolution of the commutator one has to repeat the measurement for all times.
	The result is depicted in Fig.~\ref{Fig:final22} for $2\times 10^4$ random numbers per data point and time.
	For each point in time we repeat the step which is depicted in Fig.~\ref{Fig:final11}.
	Afterward we fit the data and extract the quadratic slope.
	The commutator of interest, i.e., the theoretical prediction, can be analytically derived and reads as
	\begin{align}
	\label{Eq:JCcommutator}
	 &\frac{i}{ \hbar} \int_{0}^t d\tau_1  \langle 1,\alpha_0 |  \left[  \hat H_{\text{int},I}  (\tau_1) , \hat H_{\text{int},I}(t) \right] |1,\alpha_0 \rangle \\
	 &=  \frac{2 |\kappa|^2 \hbar}{\Delta \omega} (1-\cos \Delta \omega t ) \sum_{n=0}^\infty | f_k(n;\eta)|^2  \frac{|\alpha_0|^{2(n+k)} }{n!} e^{-|\alpha_0|^{2} } ,\nonumber 
	 \end{align}
	which is a harmonic oscillation in time.
	This result is given as the black lines in Fig.~\ref{Fig:final22}.
	
	The magenta and green dots correspond to the excitation to the second ($k=2$) and the zeroth ($k=0$) sideband, respectively.
	The results derived from the generated data resemble the theoretical results sufficiently well.
	It is noteworthy that, for an explicitly time-dependent Hamiltonian, to certify clear experimental evidence of the relevance of the non-equal-time commutators of the interaction Hamiltonian for the system dynamics, it is sufficient to demonstrate statistically significant nonzero contributions in Fig.~\ref{Fig:final22}. 
	Here we show that such a certification is, by the techniques proposed here, rather easy to do.
	
\section{Summary and Conclusions} 

	 To our best knowledge, presently no proposal of a method for the experimental verification of the non-equal-time commutators of interaction Hamiltonians does exist. 
  	 A reason for this is, that usually it is preferred to operate a certain dynamics under perfect resonance conditions. 
  	 However, the general situation with an explicitly time-dependent interaction needs to be fully understood. 
  	 The present paper aims to contribute significantly to this fundamental issue. 
   	For this purpose, we derived analytical expressions for measurable quantities, which render it possible to experimentally access the quantities of interest.

	We have shown that, for the vibronic dynamics of a laser-driven trapped ion in the resolved-sideband regime, the measurement of the electronic-state occupation probability yields the temporal evolution of the expectation value of the interaction Hamiltonian.
	From this value one can derive both the expectation value of the interaction Hamiltonian in the interaction picture and the partly integrated non-equal-time commutator of the interaction Hamiltonian.
	Statistically generated data points are only used to demonstrate that the proposed methods will work under realistic experimental conditions.
	The obtained results well approximate the analytically derived ones. 
	Thus, the detuned nonlinear Jaynes-Cummings Hamiltonian under study is appropriate to access the fundamentals of explicitly time-dependent temporal evolutions of quantum systems.
	
	For the determination of the Hamiltonian we have considered an input motional Fock state and obtained the interaction Hamiltonian in the Fock basis for the quasi-resonant excitation of the zeroth motional sideband.
	In addition, the non-equal-time commutator, which explicitly accounts for time-ordering corrections, has been investigated.
	For an initially prepared motional coherent state, the evolution of the partly time-integrated commutator can be determined. 
	This allows one to directly visualize in experiments the noncommutativity of the interaction Hamiltonian at different times.
	Our approach paves the way to study the explicitly time-dependent dynamics also for other quantum systems of interest. 
	However, this requires the reformulation of the corresponding measurement principles for the systems to be studied, which is beyond the scope of the present paper.

\end{document}